\documentstyle[aps,prl,epsfig,twocolumn]{revtex}

\def\be{\begin{equation}}
\def\ee{\end{equation}}
\def\bea{\begin{eqnarray}}
\def\eea{\end{eqnarray}}
\def\bma{\begin{mathletters}}
\def\ema{\end{mathletters}}

\begin{document}
\draft

\title{Spin monopoles with Bose--Einstein condenstates}

\author{J.J. Garc\'{\i}a--Ripoll,$^{1}$ J.I. Cirac,$^{2}$ J. Anglin,$^{2}$
V. P\'erez--Garc\'{\i}a,$^{1}$ and P. Zoller$^{2}$}

\address{
$^{1}$Departamento de M\'atem\'aticas, Universidad de Castilla--La Mancha,
13071 Ciudad Real, Spain.\\
$^{2}$Institut f\"ur Theoretische Physik, Universit\"at Innsbruck,
Technikerstrasse 25, A--6020 Innsbruck, Austria.
}

\date{\today}

\maketitle

\begin{abstract}
We study the feasibility of preparing a Bose--Einstein condensed sample
of atoms in a 2D spin monopole. In this state, the atomic internal spins 
lie in the $x$--$y$ plane, and point in the radial direction.
\end{abstract}



\narrowtext


After the successful generation of Bose--Einstein condensation of alkali
atoms, the creation and manipulation of certain macroscopic quantum
states remains as one of the fundamental goals in the field of Atomic
Physics \cite{bechomepage}. During the last year, several ways to create vortices and
solitons have been proposed \cite{Dum,Walls}. 
Under certain circumstances these states
are stable \cite{Rokhsar,Stringari,Fetter}, which has motivated several experimental groups to try
corresponding experiments. In this letter we study a new kind of
macroscopic quantum state for these atomic samples, what we call a {\em 2D
spin monopole}. It is a state in which the atomic spin (i.e., the
magnetization vector) points in the radial direction in the $x$--$y$
plane [Fig.\ 1(a)]. We show that this state is stable under realistic
conditions, and analyze a method to generate it which only requires
current experimental technology. We will first study a one dimensional
situation, in which the condensate is confined in a ring, and then we
will generalize it to the 3D case.

\begin{figure}[tbp]
\epsfig{file=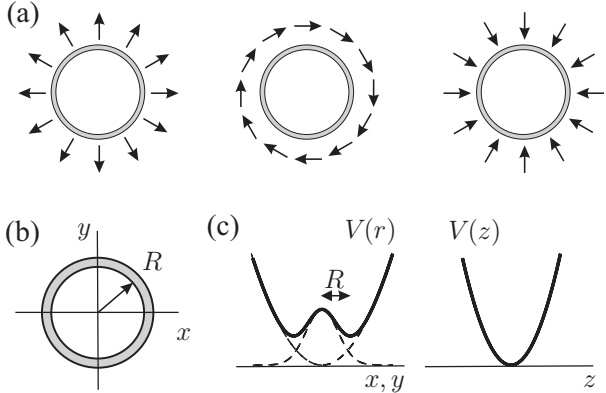,width=8cm}
\caption{(a) Spin monopole in 2D. For $\mu\ne 0$ the spins rotate as
a function of time; (b) Ring trap; (c) Potential $V(r,z)$}
\end{figure}

We consider a Bose--Einstein condensed sample of $N$ atoms. The atoms have
two internal levels, $|\uparrow\rangle$ and $|\downarrow\rangle$, which
are relevant for the dynamics \cite{Ho1,JILA}. We assume that half of the atoms is in
each internal level. The atomic motion is confined by an external trap
with a ring shape [see Fig.\ 1(b)] \cite{MIT}. In the limit where the motion along
the radial ($r$) and axial ($z$) direction is frozen, the dynamics of the
motional state only depends on the polar angle ($\theta$). We can write
the wavefunction of the condensate as
\be
\label{Psi}
|\Psi(\theta,\tau)\rangle = \phi_1(\theta,\tau) |\uparrow\rangle
 + \phi_2(\theta,\tau) |\downarrow\rangle,
\ee
where $\phi_{1,2}$ are the motional wavefunctions corresponding to the 
internal states $|\uparrow,\downarrow\rangle$, respectively and fulfill the
coupled Gross--Pitaevskii Equations:
\bma
\label{GPring}
\bea
i \frac{d}{d\tau} \phi_1 &=& \left[
 - \frac{d^2}{d\theta^2} + \delta_1 +  u_{11} |\phi_1|^2 +
  u_{21} |\phi_2|^2 \right] \phi_1,\\
i \frac{d}{d\tau} \phi_2 &=& \left[
 - \frac{d^2}{d\theta^2} + \delta_2 +  u_{12} |\phi_1|^2 +
  u_{22} |\phi_2|^2 \right] \phi_2,
\eea
\ema
with normalization
\be
\frac{1}{2\pi} \int_0^{2\pi} d\theta \; |\phi_{1,2}(\theta,\tau)|^2 = 1.
\ee
Here, $\delta_{1,2}$ denotes the energy of the two internal states and
$u_{ij}=u_{ji}$ ($i,j=1,2$) describe their mutual interactions. All the
quantities in Eqs.\ (\ref{GPring}) have been rescaled so that they are
dimensionless. In particular, the $u_{ij}$ are proportional to the
number of atoms $N$, the corresponding scattering length, and the square of 
the ring radius [see Eq.\ (\ref{uij}) below]. 

The state (\ref{Psi}) with 
\be
\label{monopole}
\phi^{\rm mp}_1(\theta,\tau) = e^{-i \mu_1 \tau},\quad
\phi^{\rm mp}_2(\theta,\tau) = e^{i\theta} e^{-i \mu_2 \tau},
\ee
is a stationary solution of Eqs.\ (\ref{GPring}) with $\mu_j = j-1+
u_{1j} + u_{2j} + \delta_j$ ($j=1,2$). Furthermore, defining the Pauli
operator $\vec\sigma=(\sigma_x,\sigma_y,\sigma_z)$ as usual, we have
that the state (\ref{Psi}) with (\ref{monopole}) is an eigenstate of
$\vec \sigma \cdot \vec n$ with $\vec n=[\cos(\theta-\mu
\tau),\sin(\theta-\mu \tau),0]$ and $\mu\equiv\mu_2-\mu_1$. This means
that for $\mu=0$ the spins are always pointing outwards in the $x$--$y$
plane, whereas for $\mu\ne 0$ they oscillate as shown in Fig.\ 1(a).
Thus, the stationary state (\ref{monopole}) can be considered as an
oscillating spin monopole in two dimensions. 

In order to analyze the stability of the spin monopole we have
carried out a perturbative linear analysis. We consider a small
perturbation around the monopole solution, such that
$\phi_{1,2}(\theta,\tau)=\phi_{1,2}^{\rm mp}(\theta,\tau) + \epsilon
\alpha_{1,2}(\theta,\tau)$. We then expand Eq.\ (\ref{GPring}) up to
first order in $\epsilon$ obtaining a linear set of coupled equations
for $\alpha_{1,2}$ and $\alpha_{1,2}^\ast$. Expanding
$\alpha_{1,2}(\theta,\tau)= \sum_{n=-\infty}^{\infty}
\alpha_{1,2}^{(n)}(\tau) e^{in\theta}$ and substituting in these
equations, we obtain
\be
\label{stability}
i\frac{d}{d\tau} \vec \alpha^{(n)} = E {\cal H}_n
  \vec\alpha^{(n)},
\ee
where
\be
\vec \alpha^{(n)} =
  \left[\alpha_1^{(n)},\alpha_1^{(-n)\ast},\alpha_2^{(n+1)},
  \alpha_2^{(-n+1)\ast}\right]^{\rm T},
\ee
is a column vector. Here, ${\cal H}_n \equiv K_n + H^{\rm int}$, with
\bma
\bea
E &=& {\rm diag}(1,-1,1,-1),\\
K_n &=& {\rm diag}[n^2,n^2,(n+1)^2-1,(n-1)^2-1],\\
H^{\rm int} &=& 
 \left( \begin{array}{cc} u_{11} & u_{21} \\ u_{12} & u_{22} \end{array}\right) 
 \otimes
 \left( \begin{array}{cc} 1 & 1 \\ 1 & 1 \end{array} \right),
\eea
\ema
are $4\times 4$ matrices and $\otimes$ denotes tensor product. The
stability analysis can be carried out by diagonalizing the matrices $E
{\cal H}_n$. Complex eigenvalues $\pm i\lambda$ correspond to
exponentially growing solutions [$\alpha(\tau)=\alpha(0) e^{\lambda
\tau}$], i.e. the stationary solution is {\rm dynamically unstable}.
Real eigenvalues $\lambda$ lead to small oscillatory solutions
[$\alpha(\tau)=\alpha(0) e^{\pm i \lambda \tau}$], which may correspond
to higher or lower values of the Gross--Pitaevskii energy functional
with respect to the stationary solution. In the first case, the
stationary solution is {\rm stable}, whereas in the latter one the
system is {\rm energetically unstable}; this means that if energy can be
taken out from the system at a given rate, then the stationary solution
is unstable on that time scale. Another interesting case occurs when one
of the matrices is not simple (i.e., non diagonalizable), since in that
case one has solutions which only grow polynomically with time [for
example, for a $2\times 2$ Jordan matrix corresponding to a degenerate
eigenvalue $\lambda$, $\alpha(\tau)=\alpha(0) (1+\lambda \tau)$]. On the
other hand, if one is only interested in checking whether the solution
is stable or not, (without distinguishing among the different types of
instability) one can simply study the positivity of ${\cal H}_n$. If
all matrices ${\cal H}_n\ge 0$ and the projector operator $P_0^n$ on the
kernel of ${\cal H}_n$ commutes with $E$ for all $n$, then the stationary
solution is stable. One can proof this statement as follows: first, the
positivity of ${\cal H}_n$ ensures the positivity of the Gross--Pitaevskii
energy functional and therefore there cannot be energetically
instabilities; secondly, all eigenvalues of $E {\cal H}_n$ are also
eigenvalues of ${\cal H}_n^{1/2} E {\cal H}_n^{1/2}$, which is hermitian
(has real eigenvalues) and therefore there cannot be dynamical
instabilities; finally, if $[P_0^n,E]=0$ one can easily prove that if
a vector belongs to the kernel (range) of ${\cal H}_n$, then it also
belongs to the kernel (range) of $E{\cal H}_n$, and therefore this last
matrix is simple.

We start out by analyzing the positivity of ${\cal H}_n$. We consider
two cases: (a) $u_{11} u_{22} < u_{12}^2$. The matrix $H^{\rm int}$ has
a negative eigenvalue; for $n=0$ we have $K_0=0$ and therefore ${\cal
H}_0$ has a negative eigenvalue and the solution is not stable. (b)
$u_{11} u_{22} > u_{12}^2$. Now $H^{\rm int} > 0$. Given that $K_n\ge 0$
for $|n|>2$ we can restrict ourselves to the case $|n|=1$. Since the
first three diagonal minors of ${\cal H}_n$ are positive, we just have
to impose that the determinant be positive. We obtain that the monopole
(\ref{monopole}) is stable if $u_{11} u_{22} - u_{12}^2 + u_{22}/2 - 3
u_{11}/2 - 3/4 > 0$. Now, we investigate which kind of instabilities
occur in the opposite case. We can again restrict ourselves to the case
$|n|=1$, and therefore we have to find the eigenvalues $\lambda$ of the
matrix $E {\cal H}_1$. In the relevant case for which
$u_{11}=u_{22}\equiv u$ (see below) this can be done analytically. We
obtain:
\be
\label{lambda}
\lambda= 1\pm \left[2\left(1+u\pm\sqrt{1+2u+u_{12}^2}\right)\right]^{1/2}.
\ee
Thus, we have the following scenario:
\[
\begin{array}{rc}
u_{12}^2 \le u^2-u-3/4 \quad     & \quad {\mbox{\rm Stable}} \\
u^2-u-3/4 < u_{12}^2 < u^2 \quad & \quad {\mbox{\rm Energetically unstable}} \\
u^2 < u_{12}^2           \quad   & \quad {\mbox{\rm Dynamically unstable.}} 
\end{array}
\]
The eigenvalues as well as the stability diagram is shown in Fig.\ 2.
For $u>u_{12}$ and when the interaction energy becomes more important
than the (rotational) kinetic energy ($u\gg 1$), the solution is stable as 
long as $u-u_{12} \agt 1$.
Since both $u$ and $u_{12}$ are proportional to the number of atoms $N$, we have
that by increasing $N$ one can completely stabilize the spin monopole. An
interesting situation occurs for $u=u_{12}$; in that case the matrix $E
{\cal H}_1$ is not simple, which implies that the perturbation only
grows linearly with time. On the other hand, the dynamical instability
occurring for $u<u_{12}$ also appears for the homogeneous stationary
solution and therefore it simply corresponds to a phase separation of
the two components (internal states), as could be expected. 

\begin{figure}[tbp]
\epsfig{file=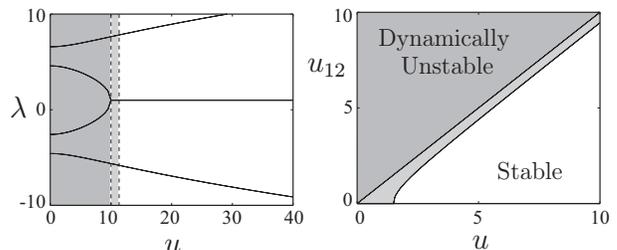,width=8cm}
\caption{(a) Excitation frequencies as a function of $u$ for $u_{12}=10$;
(b) Stability diagram}
\end{figure}

In order to establish if the monopole continues being stable in a
realistic 3--dimensional situation, we consider a trapping potential
of the form [Fig.\ 1(c)]
\be
V(r,z) = \frac{1}{2}m\omega_z^2 z^2 + \frac{1}{2}m\omega_r^2 r^2
+ V_0 e^{-r^2/(2\sigma^2)}.
\ee
This corresponds to a standard dipole trap with an off resonant Gaussian
laser beam propagating along the $z$ direction \cite{MIT}. Here $V_0$ gives the
ac--Stark shift at the center of the trap. The equilibrium point $R$
along the radial direction is given by $R=2\sigma^2
\ln[V_0/(m\omega_r^2\sigma^2)]$. In order to have a ring trap, we choose
the parameters such that $R\gg \Delta r$, where $\Delta r$ is the
typical size of the atomic cloud along the $z$ and $r$ directions.
Writing $|\Psi(r,z,\theta)\rangle = \psi_1(r,z,\theta) |\uparrow\rangle
+ \psi_2(r,z,\theta) |\downarrow\rangle$, the Gross--Pitaevskii Equation
governing the evolution is 
\bea
i \hbar \frac{d}{dt} \psi_1 &=& \left[ - \frac{\hbar^2\nabla^2}{2m} + 
 V + \hbar\tilde\delta_1 + \tilde u_{11} |\psi_1|^2 + \tilde u_{12} |\psi_2|^2 
 \right] \psi_1,\nonumber \\
i \hbar \frac{d}{dt} \psi_2 &=& \left[ - \frac{\hbar^2\nabla^2}{2m} + 
 V + \hbar\tilde\delta_2 + \tilde u_{21} |\psi_1|^2 + \tilde u_{22} |\psi_2|^2 
 \right] \psi_2.\nonumber
\eea
Here $\tilde u_{ij}=4\pi\hbar^2 a_{ij}/m$ with $a_{ij}=a_{ji}$ being the
s--wave scattering lengths corresponding to the different collisions,
and $\tilde\delta_{1,2}$ two constant offsets. The functions
$\psi_{1,2}$ are normalized to the number of atoms $N_{1,2}$ in each
internal state.

Let us first make the connection between the 3--D model and the ring. We
consider the simple situation in which the motions along the radial and
$z$ direction are frozen. The conditions of validity of this situation
will be discussed below. If the number of atoms in each internal level
is the same $N_1=N_2\equiv N/2$, we can reduce the full three
dimensional problem to the ring case studied above by writing
$\psi_{1,2}(r,z,\theta,t)=[N/(4\pi)]^{1/2}f_{1,2}(r,z)\phi_{1,2}(\theta,t)$,
multiplying the coupled Gross--Pitaevskii equations by $f_i^\ast$ and
integrating in $r$ and $z$. We obtain that $\phi_{1,2}$ satisfy the
Eqs.\ (\ref{GPring}) with
\bma
\bea
\label{uij}
u_{ij} &=& 2N R^2 a_{ij}  \int_{-\infty}^{\infty} dz
\int_0^\infty rdr |f_i(r,z)|^2 \; |f_j(r,z)|^2,\\
\delta_i &=& \frac{2mR^2}{\hbar^2} (\hbar\tilde \delta_i + \epsilon_i),\quad
\tau =  \frac{\hbar}{2mR^2} t,
\eea
\ema
where $\epsilon_i$ is the expectation value of the kinetic plus potential
energy with the wavefunction $f_i$. 

In general, we look for stationary solutions of the form
\bma
\label{mono3D}
\bea
\psi_1(r,z,\theta,t) &=& \sqrt{N/2} f_1(r,z)e^{-i \tilde\mu_1 t},\\
\psi_2(r,z,\theta,t) &=& \sqrt{N/2} f_2(r,z)e^{i\theta} e^{-i \tilde\mu_2 t},
\eea
\ema
with $f_{1,2}$ normalized real functions satisfying $\tilde L_{1,2} f_{1,2} = 0$
where
\bea
\label{L}
\tilde L_{n} &=& -\frac{\hbar^2}{2m} \left[ \frac{d^2}{dr^2} + 
  \frac{1}{r}\frac{d}{dr} + 
  \frac{d^2}{dz^2} - \frac{(n-1)^2}{r^2}\right] + V(r,z) \nonumber \\
&& - \hbar\tilde \delta_n - \hbar\mu_n
  + \tilde u_{1n} f_1^2 + \tilde u_{2n} f_2^2.
\eea
We will concentrate in the case $\tilde u_{11}=\tilde u_{22} \equiv
\tilde u$ (equivalently, $a_{11}=a_{22}\equiv a$) which is the relevant
experimental situation described below. In the limit $R\gg \Delta r$ the
centrifugal term in $\tilde L_2$ can be approximated by a constant
$\hbar^2/(2mR^2)$. If there is no phase separation ($a >
a_{12}$) we find $f_1(r,z) = f_2(r,z) \equiv f(r,z)$ and $\tilde
\mu\equiv \tilde\mu_2-\tilde\mu_1= \hbar/(2mR^2) +\tilde\delta_2
-\tilde \delta_1$. Solution (\ref{mono3D}) is again an eigenstate of $\vec n
\cdot \vec \sigma$ with $\vec n=[\cos(\theta-\tilde\mu
t),\sin(\theta-\tilde\mu t),0]$, being the eigenvalue proportional to
$|f(r,z)|^2$. Thus it represents a state where the atoms have their spin
in the $x$-$y$ plane forming a 2D monopole. The length of the spin at
each point depends on the corresponding local density. We analyze now
the stability of such a solution. As before, we linearize around the
solution (\ref{mono3D}) by adding a small quantity $\alpha_{1,2}$ and
expanding it in powers of $e^{i\theta}$. We obtain (\ref{stability}) where
now ${\cal H}_n= \tilde L + \tilde K_n + \tilde H^{\rm int}$ a $4\times
4$ matrix operator with $\tilde L={\rm diag}(\tilde L_1,\tilde
L_1,\tilde L_2,\tilde L_2)$, $\tilde K_n=\hbar^2/(2mr^2) K_n$, and
$\tilde H^{\rm int} = f^2 H^{\rm int}$. Using again $R\gg \Delta r$ we
have $L_1=L_2$ and therefore $\tilde L=L_1$ times the $4\times 4$
identity matrix. Since we are only interested in the stability, we 
analyze the positivity of ${\cal H}_n$. As before, we just have to
study the cases $|n|<2$. Let us distinguish two situations:

{\em Weak interactions}: in the limit $N(a+a_{12})/R \ll 1$ we have that
the interaction energy $N\tilde u/[2\pi R (\Delta r)^2]$ is much smaller
than the harmonic oscillator quantum $\hbar\omega$. In that case,
$\Delta r\simeq a_0$ where $a_0=[\hbar/(m\omega)]^{1/2}$ is the size of
harmonic potential ground state. The spectrum of ${\cal H}_{0,1}$ is
dominated by $\tilde L$. The radial and $z$ dependence give rise to
excitation energies $k\hbar \omega$ ($k$ integer). The lowest
excitations $k=0$ correspond to $\alpha(r,z) \propto f(r,z)$, and
therefore we obtain $\tilde \lambda=\hbar^2/(2mR^2) \lambda$ ($\ll
\hbar\omega$ in absolute value), where $\lambda$ is given in
(\ref{lambda}) with $u$ and $u_{12}$ given in (\ref{uij}). Thus, for
excitation energies lower than $\hbar\omega$ the problem fully reduces
to the ring case.

{\em Strong interactions}: In the opposite limit, we are in the
Thomas--Fermi regime, where $\Delta r=a_0 [32 N(a+a_{12})/R]^{1/4}$.
Now, one cannot simply separate radial and $z$ excitations from ring
excitations. The excitation spectrum of ${\cal H}_{0,1}$ is dominated by
$\tilde L + \tilde H^{\rm int}$. It is convenient to diagonalize $\tilde
H^{\rm int}$, and consider the eigenfunctions separately. (a) Consider
$\vec\alpha^{(n)}=(g_1,g_2)\otimes (1,-1)$: in this case $\tilde H^{\rm
int}$ is zero, and therefore the excitation frequencies correspond to
those of $\tilde L$, which are of the order of $k \hbar^2/[2m(\Delta
r)^2] \gg \hbar^2/(2mR^2)$. (b) For $\vec\alpha^{(n)}=(g,g)\otimes
(1,1)$, $H^{\rm int}$ gives $(\tilde u + \tilde u_{12})f^2$, whereas for
$\vec\alpha^{(n)}=(g,-g)\otimes (1,1)$ it gives $(\tilde u - \tilde
u_{12})f^2$: the lowest energy will be of the order of $N(\tilde u
-\tilde u_{12})/[2\pi R (\Delta r)^2]$. As long as this energy is larger
than $\hbar^2/(2mR^2)$ we can consider separately the cases (a) and (b)
treating $K_n$ as a perturbation; in both the correction is positive,
i.e. the monopole is stable. In the opposite case, one has to be more
careful in the perturbation analysis, since one cannot separate the
cases (a) and (b); the excitation energies may become negative. Thus, we
obtain a necessary condition for stability $N(\tilde u -\tilde
u_{12})/[2\pi R (\Delta r)^2]\gg \hbar^2/(2mR^2)$. Using (\ref{uij}) we
can write this condition as $u-u_{12}>1$, which coincides with the basic
stability condition derived for the ring. It means that the interactions
have to be sufficiently strong to stabilize the monopole.

In order to be specific, we will propose now a particular configuration
to create the spin monopole. We consider an alkali atom in a ground
$F=1$ hyperfine state. We will assume that the energy of the $m_F=0$
level is made higher (by using an off--resonant laser or radio-frequency
field), so that it is not involved in the dynamics. In this case we can
identify $|\uparrow\rangle =|F=1,m_F=1\rangle$ and $|\downarrow\rangle
=|F=1,m_F=-1\rangle$, the collisions do not change spin, and $\tilde
u_{11}=\tilde u_{22}\equiv \tilde u$. Although the current status does
not allow to make a clear statement about the sign of $\tilde u-\tilde
u_{12}$ it seems that for $^{23}Na$ it is positive, which is also agrees
with recent experimental results concerning miscibility \cite{Ho2}. In
order to generate the 2--D spin monopole (\ref{mono3D}) we propose to
use an off-resonant Raman beam. The atoms are initially condensed in the
internal $|\uparrow\rangle$ state. A Raman laser that connects the
states $|\uparrow\rangle$ and $|\downarrow\rangle$ is then switched on.
It should have the appropriate spatial dependence so that the angular
momentum in the $z$ direction is changed by one unit \cite{note}.
Denoting by $\Omega(z,r,\theta)$ the effective Rabi frequency, the
evolution equations are the above Gross--Pitaevskii Equations but with a
coupling term between $\psi_1$ and $\psi_2$ proportional to $\Omega$;
the Raman detuning is incorporated to the definitions of $\tilde
\delta_{1,2}$. Initially, one takes $\tilde\delta_1\gg \tilde\delta_2$,
so that the laser does not affect the internal atomic state, since it is
effectively out of resonance. Then, $\tilde \delta_2-\tilde\delta_1$ is
changed adiabatically until $\tilde\delta_2-\tilde\delta_1\simeq 0$. The
method is more robust than the one used to generate vortices since the
spatial wavefunctions $f_{1,2}(r,z)$ remain practically constant with
our setup. Actually, in this case one can simply use a $\pi/2$ laser
pulse, taking $\tilde\delta_1=\tilde\delta_2$. This allows to generate
the monopole state in a much faster time scale, which will be of the
order of several inverse trap frequencies.

\begin{figure}[tbp]
\epsfig{file=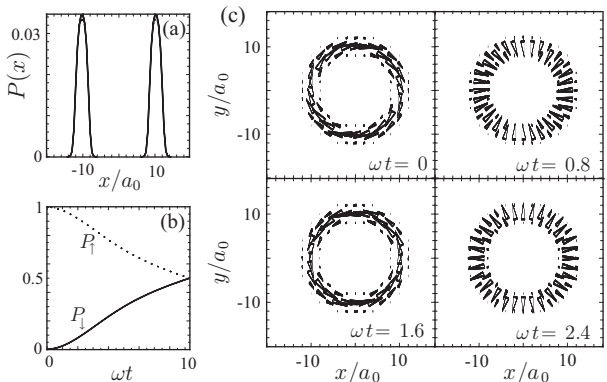,width=8cm}
\caption{Preparation of the 2D spin monopole. Trap parameters: $\omega_z=2\omega_r$,
$R=10a_0$, $\sigma=5a_0$, $V_0=200 \hbar\omega_r$, where $a_0=(\hbar/m\omega)^{1/2}$;
interactions: $\tilde u_{11}=\tilde u_{22}=\tilde u_{12}/0.9 = 3600 \hbar\omega_r a_0^3$;
Laser parameters: $\Omega(r)=\Omega_0[\sin(kx)+i\sin(ky)]$ with $\Omega_0=0.28\omega_r$
and $k=\pi/(6R)$. (a) Final density distribution $|\psi_1|^2$ (dashed line) and
$|\psi_2|^2$ (solid line) at $y=z=0$ as a function of $x$; (b) Evolution of
the population of the $|\uparrow\rangle$ (dashed line) and $|\downarrow\rangle$
(solid line); (c) Evolution of the spin density after the preparation for taking 
$\delta_2-delta_1=2\omega_r$; the triangles point along the expectation value
of $\langle \sigma\rangle$ and are proportional to the local density.
}
\end{figure}

In order to evaluate our proposal, we have performed a 3--dimensional
numerical simulation of the Gross--Pitaevskii equations in the presence
of the laser for the creation of the spin monopole. We have used an
optimized three-dimensional collocation Fourier method with typically
80$\times$80$\times$40 collocation points and integrating in time with a
symmetrized split-step operator technique. The results are shown in
Fig.\ 3. In Fig. 3(a) we have plotted the final density profile along
the $x$ axis at the end of the generation process, whereas in Fig.\ 3(b)
we have plotted the evolution of the population of the internal levels
until the monopole is generated. After this process, we switch of the
laser and apply an internal energy shift so that the spin start
precessing, as is shown in Fig.\ 3(c). For this figure we have taken
realistic parameters. For example, taking Na with $\omega_r=100$Hz,
$a_0=2\mu$m, and $a=52a_B$ we have that the number of atoms is of the
order of $2\times10^5$. In order to ensure the stability, we have
evolved the formed state in imaginary time (renormalizing the state
after each evolution step) for about 20 trap oscillation times, without
noticing any instability.

In summary, we have shown that when the interactions are sufficiently
strong, a 2D spin monopole becomes stable in a ring trap. We have
performed the stability analysis, both in a ring situation as well as in
the full 3D case. We have shown a way to prepare such a state, and
verified with a full 3D numerical simulation that one can prepare
it with current experimental parameters.

J.J.G-R and V.M.P-G. have been partially supported by CYCIT grants
PB96-0534 and PB95-0839. The work in Innsbruck has been supported
by the Austrian Science Foundation.



\begin{references}
\bibitem{note}
As proposed in Ref.\ \cite{Dum} this can be achieved by using two standing waves propagating
along the $x$ and $y$ axis with a phase diference, i.e., an effective electric field
of the form $\sin(kx)+i\sin(ky) \simeq k(x+iy)$ if $k$ is sufficiently small. This
last requirement can be achieved using pairs of beams forming the appropriate
angle.
\end{references}
\end{document}